\documentclass[conference]{IEEEtran}
\IEEEoverridecommandlockouts
\usepackage{cite}
\usepackage{amsmath,amssymb,amsfonts}
\usepackage{algorithmic}
\usepackage{graphicx}
\usepackage{textcomp}
\usepackage{xcolor}
\usepackage{empheq}
\usepackage{caption}
\usepackage{subcaption}

\usepackage[a4paper, total={184mm,239mm}]{geometry}
\def\BibTeX{{\rm B\kern-.05em{\sc i\kern-.025em b}\kern-.08em
    T\kern-.1667em\lower.7ex\hbox{E}\kern-.125emX}}

\usepackage{multirow}
\usepackage{booktabs}

\usepackage{footnote}
\makesavenoteenv{tabular}

\newcommand{\blockBegin}[1][c]{\begin{array}{#1}}
\newcommand{\blockEnd}{\end{array}}
\newcommand{\curlyBegin}[1][c]{\left \{ \!\!\! \begin{array}{#1}}
\newcommand{\curlyEnd}{\end{array} \!\!\!\!\! \right\}}
\newcommand{\ceilBegin}[1][c]{\left \lceil \!\!\! \begin{array}{#1}}
\newcommand{\ceilEnd}{\end{array} \!\!\! \right\rceil}

\newcounter{daggerfootnote}

\begin{document}

\title{
Monomorphism-based CGRA Mapping \\via Space and Time Decoupling 
\thanks{This work was supported by the Swiss National Science Foundation via project ADApprox (grant 200020\_188613).}
}

\author{
\IEEEauthorblockN{Cristian Tirelli}
\IEEEauthorblockA{\textit{Faculty of Informatics} \\
\textit{Universit{\`a} della Svizzera italiana} \\
Lugano, Switzerland \\
cristian.tirelli@usi.ch}
\and
\IEEEauthorblockN{Rodrigo Otoni}
\IEEEauthorblockA{\textit{Faculty of Informatics} \\
\textit{Universit{\`a} della Svizzera italiana} \\
Lugano, Switzerland \\
otonir@usi.ch}
\and
\IEEEauthorblockN{Laura Pozzi}
\IEEEauthorblockA{\textit{Faculty of Informatics} \\
\textit{Universit{\`a} della Svizzera italiana} \\
Lugano, Switzerland \\
laura.pozzi@usi.ch}
}

\maketitle


\begin{abstract}
Coarse-Grain Reconfigurable Arrays (CGRAs) provide flexibility and energy efficiency in accelerating compute-intensive loops. Existing compilation techniques often struggle with scalability, unable to map code onto large CGRAs. To address this, we propose a novel approach to the mapping problem where the time and space dimensions are decoupled and explored separately. We leverage an SMT formulation to traverse the time dimension first, and then perform a monomorphism-based search to find a valid spatial solution. Experimental results show that our approach achieves the same mapping quality of state-of-the-art techniques while significantly reducing compilation time, with this reduction being particularly tangible when compiling for large CGRAs. We achieve approximately $10^5\times$ average compilation speedup for the benchmarks evaluated on a $20\times 20$ CGRA.
\end{abstract}



\section{Introduction}

\begin{figure}[b]
    \centering
    \includegraphics[width=0.98\linewidth]{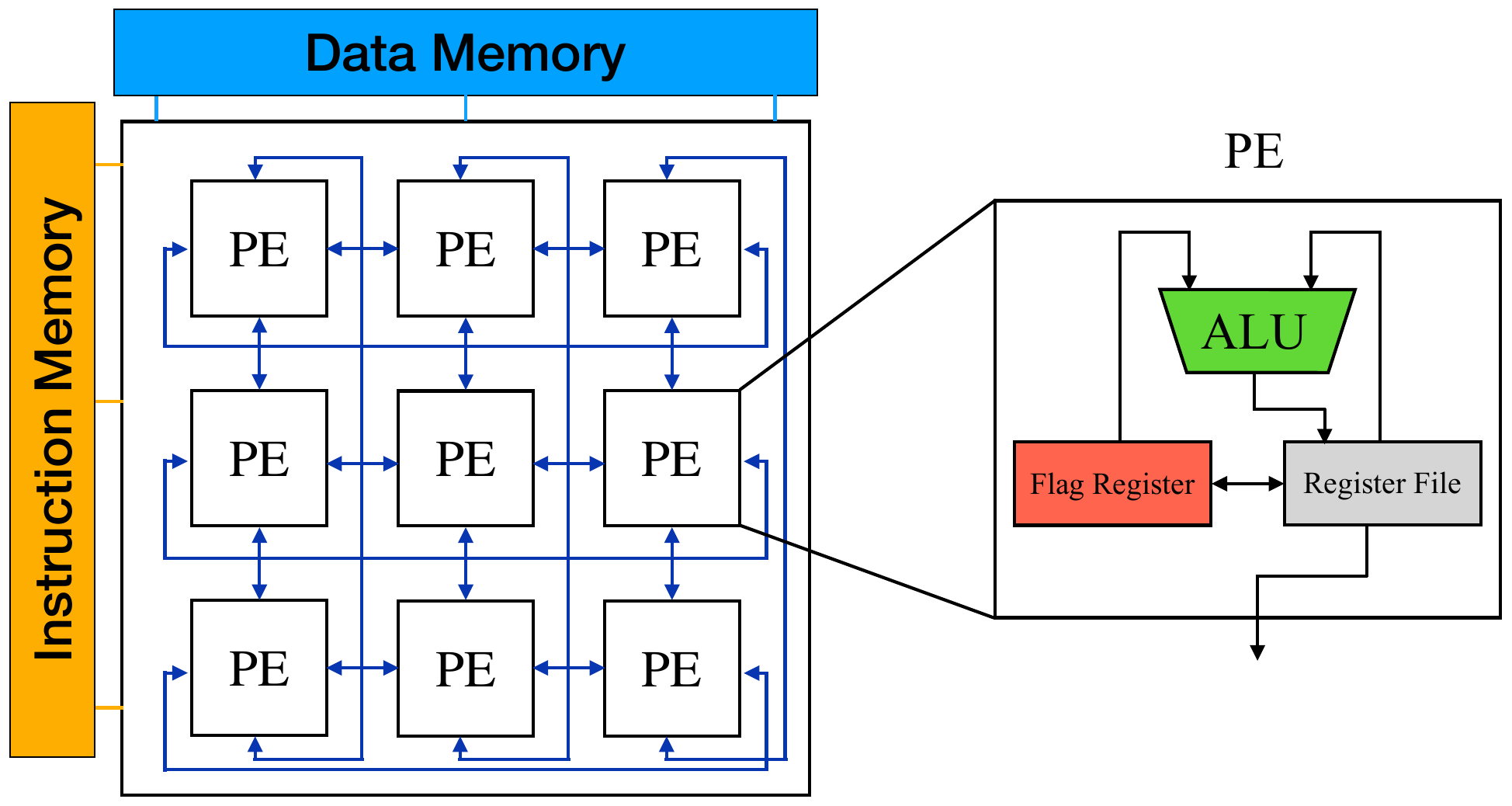}
    \caption{$3\times 3$ CGRA with internal view of a PE.}
    \label{fig:cgra-arch}
\end{figure}

As everyday applications require evermore computational power, there has been a growing need for high-performance and low-power architectures. Various solutions exist to efficiently perform compute-intensive tasks under tight power-resource constraints, some more effective than others.
Application Specific Integrated Circuits~(ASICs) accelerators, for example, provide excellent energy performance, but they are limited by their fixed functionality. 
Field Programmable Gate Arrays~(FPGAs) are more flexible, since they can be reconfigured for different applications as needed, being ideal for prototyping and custom hardware implementations; this flexibility, however, comes at the cost of lower energy efficiency~\cite{lin2014finegrain, kuon2006measuring}.
Coarse-Grain Reconfigurable Arrays~(CGRAs) offer a balanced compromise, with run-time reconfigurability at the instruction level and high computational efficiency~\cite{li2021chordmap, karunaratne2017hycube}. These characteristics are well suited to domains such as streaming and multimedia applications~\cite{akbari2018px, oh2009recurrence, lee2015optimizing, wijerathne2019cascade, duch2017heal}, but also in edge domains where they enable flexible hardware acceleration in resource-constrained scenarios.
A CGRA is composed of a set of Processing Elements~(PEs) usually organized in a 2D mesh near-neighbor topology. Every PE contains an Arithmetic Logic Unit~(ALU) and a number of internal registers organized in a register file, as depicted in Fig.~\ref{fig:cgra-arch}. Besides being connected to their neighbors according to a mesh topology, PEs also share a connection to an external memory, in order to load inputs and store outputs.

The main challenge of CGRA exploitation is the compilation process: translating high-level code onto the CGRA while taking advantage of the parallelism of the architecture. Compilers achieve this by following three steps: identify the code loops to be accelerated, extract their Data Flow Graphs~(DFGs), and find valid \emph{space-time} mappings for them. The latter notion refers to mapping DFG instructions to the right \emph{place}, i.e., assigning them to PEs in a way that data dependencies are respected through the CGRA network, and at the right \emph{time}, i.e., scheduling DFG instructions so that the data produced by every PE is consumed at the right time by other PEs.
The quality of the space-time mapping directly affects performance,
and existing mapping techniques rely on heuristics to schedule, place, and then route operations and data on the PEs. They suffer from limited scalability, however, which hampers their practical usage. To address this, we propose and evaluate a novel monomorphism-based approach for scalable CGRA mapping. Our idea is to decouple the spatial and temporal dimensions of the mapping process. First, we find a time solution that ensures spatial feasibility, and then we use a monomorphism search algorithm to efficiently find a solution in the spatial dimension. We provide a proof that a monomorphism is always present for time solutions under our constraints.

To summarize, our contributions are the following:

\begin{enumerate}
    \item Decoupling of space and time for CGRA mapping.
    \item Monomorphism-based mapping for CGRA compilation.
    \item Proof of monomorphism presence given a time solution.
    \item Evaluation showcasing the scalability of our approach.
\end{enumerate}


\section{Related Work} \label{sec:related_works}

A recent survey~\cite{podobas2020survey} summarized the evolution of CGRA architectures and methodology advancements in the last thirty years. Here, we focus on the CGRA mapping problem, i.e., the mapping of application instructions onto PEs that is an integral part of the compilation of software source code onto a CGRA. Solutions proposed in the literature can be divided into two categories: heuristics and exact methods. 

On the heuristics category, one of the first approaches proposed used simulated annealing for scheduling, placement, and routing altogether~\cite{mei2007adres}.
Addressing all three tasks simultaneously, however, leads to long execution times, low-quality solutions, and limited scalability. 
As an alternative, an edge-centric approach 
was proposed~\cite{park2008edge}. This method generates a mapping by prioritizing the routing of each edge, with placement being handled as a consequence of the routing process.
Subsequently, leveraging graph-based techniques to solve the mapping problem became increasingly prevalent in the literature. For instance, EPImap~\cite{hamzeh2012epimap} proposed an epimorphic mapping method, where a solution is found by searching for a mapping of the DFG onto the Modulo Routing Resource Graph (MRRG)~\cite{park2006modulo}, which is a graph representation of the architecture's evolution through time.
Epimap's performance was later improved by GraphMinor~\cite{chen2014graph} and REGIMap~\cite{hamzeh2013regimap}, by reducing the mapping problem to the graph minor and maximum clique problem. REGIMap was extended to also be aware of the internal register of the PEs~\cite{dave2018ramp}.
More recently, CRIMSON~\cite{balasubramanian2020crimson} proposed a randomized iterative MS algorithm that explores the scheduling space more efficiently. An improved version of it, PathSeeker~\cite{pathseeker}, is able to analyze the mapping failures and perform some local adjustments to the schedule to obtain shorter compilation time and better solutions.
All these approaches attempt to first guess a schedule and then search for a spatial solution. They, however, do not guarantee that a spatial mapping can be found for a given time solution. In our work, we prove that if a schedule with some specific properties is identified, a valid space mapping always exists.

Exact methods try to abstract and optimally solve the mapping problem. One of the earlier works in this category proposes an integer linear programming~(ILP) formulation and proves the feasibility of mapping for a given Iteration Interval~($II$)~\cite{chin2018architecture}. Alternatively, the usage of a Boolean satisfiability~(SAT) formulation instead of an ILP one was later proposed~\cite{miyasaka2020sat}.
This work was the first effort aimed at exploiting the power of modern SAT solvers in this context, but it has a critical limitation: it is not capable of modeling loop-carried dependencies. This limitation was recently overcome by the formulation of SAT-MapIt~\cite{Tirelli2023}, which was shown to outperform comparable approaches in a hardware-agnostic context~\cite{TirelliExtended}.

EPImap\cite{hamzeh2012epimap} uses epimorphism to find a mapping, and in this sense it is related to our monomorphism method. The main differences are that it addresses the space and time dimensions simultaneously, and that it adds routing nodes to the DFG to solve the mapping problem, leading to increased $II$. An initial approach to time and space decoupling, with a temporal search similar to EPImap's method, was presented in~\cite{towardshigh}. By adding routing nodes to the DFG, however, it leads to increased $II$. A last approach of note is HiMap~\cite{wijerathne2021himap}, which uses hierarchical mapping for high scalability, but targets multidimensional kernels and is outside the scope of our work.


\section{Background} \label{sec:background}


\subsection{Compilation}

\begin{figure*}[t]
    \centering
    \begin{subfigure}{0.2\textwidth}
        \centering
        \raisebox{1.5cm}{\includegraphics[width=\textwidth]{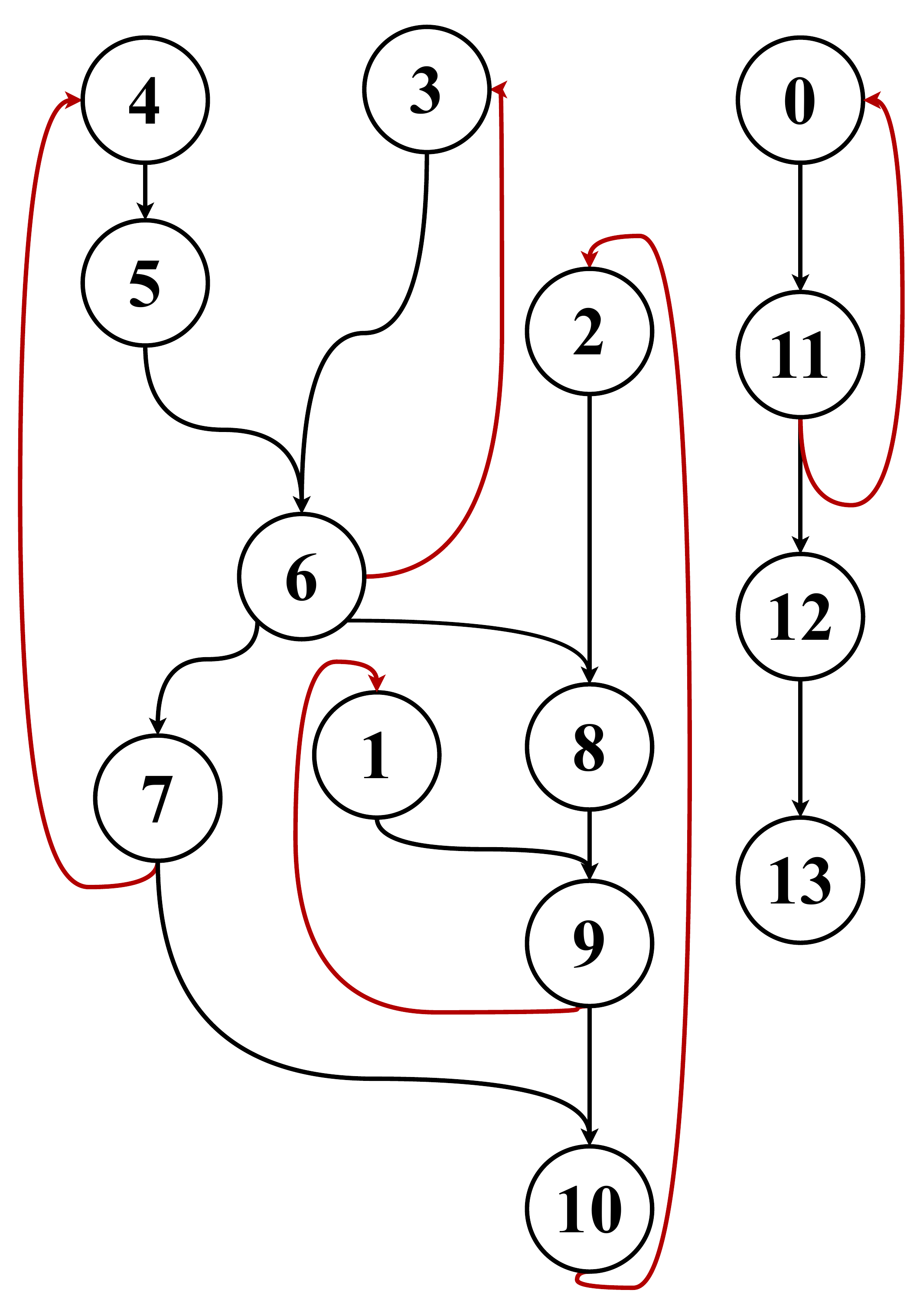}}
        \caption{}
        \label{fig:example_a}   
    \end{subfigure}\hfill
    \begin{subfigure}{0.38\textwidth}
        \centering
        \includegraphics[width=\textwidth]{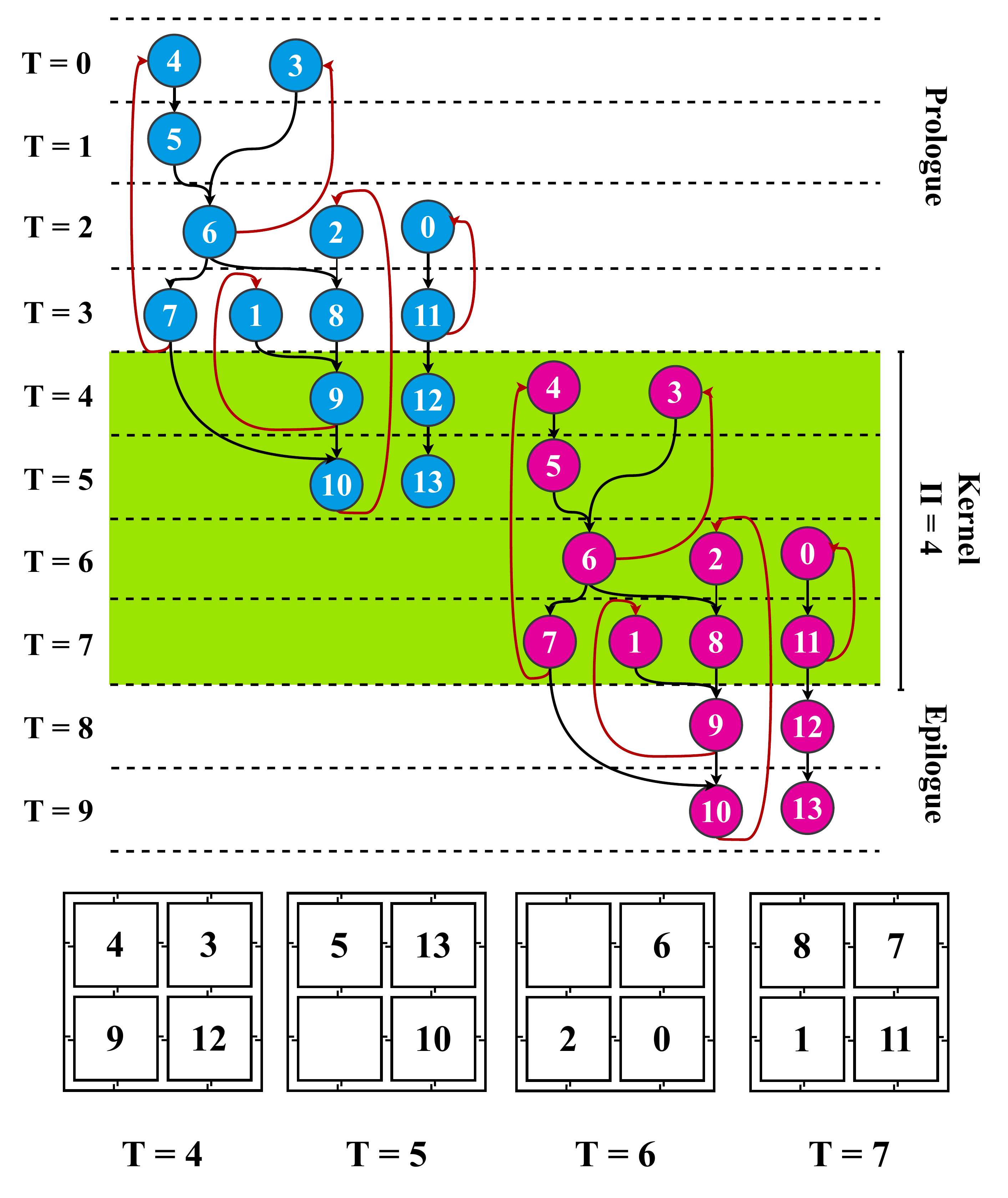}
        \caption{}
        \label{fig:example_b}   
    \end{subfigure}\hfill
    \begin{subfigure}{0.42\textwidth}
        \centering
        \raisebox{0.05cm}{\includegraphics[width=\textwidth]{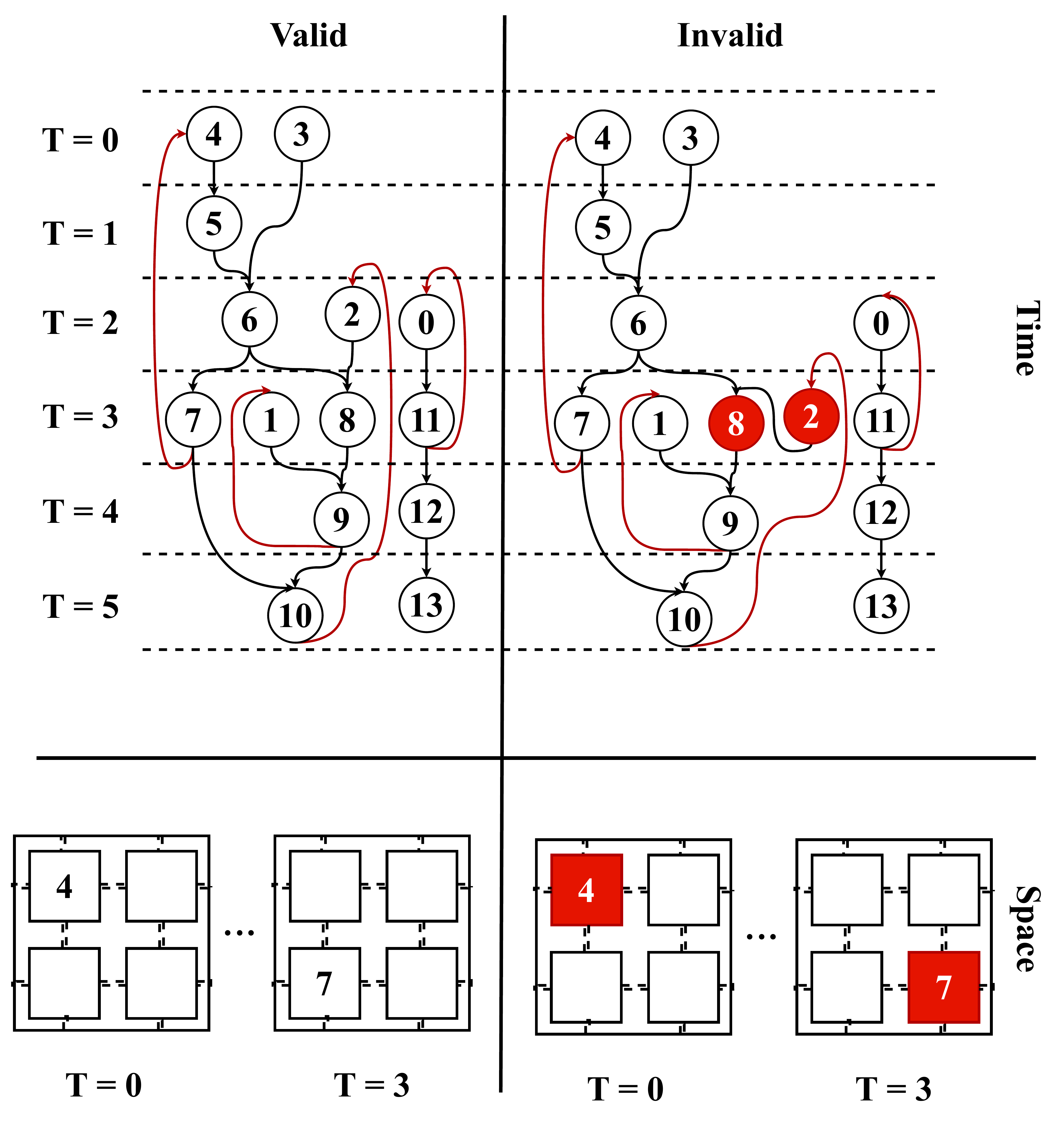}}
        \caption{}
        \label{fig:example_c}   
    \end{subfigure}
    \caption{Running example. a)~A DFG, where black edges are data dependencies and red edges are loop-carried dependencies. b)~Mapping of the DFG onto a $2\times 2$ CGRA, on the bottom, with the division between prologue, kernel, and epilogue highlighted. c)~Valid time and space solutions on the left, invalid solutions on the right; the erroneous allocations are shown in red.}
    \label{fig:example}
\end{figure*}

The choice of which compute-intensive loop should be accelerated is the first step of the compilation process. It can be done automatically~\cite{zacharopoulos2018regionseeker} or manually via pragma-annotations, with the latter being the method used in our approach. 
After the loop has been chosen, we convert it into the LLVM intermediate representation~\cite{LattnerA:2004}. From there, we extract a DFG, whose nodes represent instructions and edges represent data dependencies and loop-carried dependencies. Fig.~\ref{fig:example_a} depicts a DFG, which we will use as running example. The next step is the mapping phase, where each node of the DFG is assigned to a specific PE at a given cycle. This is where our approach is applied.

\subsection{Modulo Scheduling}

To exploit the CGRA's architectural capabilities, we make use of  Modulo Scheduling (MS)~\cite{rau1996iterative}, an optimization technique aimed at reducing the number of cycles needed to execute a loop by interleaving multiple iterations of it.
Once applied, the chosen loop is divided into three stages: prologue, kernel, and epilogue. The prologue and epilogue are executed only once: the former prepares the data for pipelining, while the latter finalizes the data produced from the pipeline. The kernel is repeated multiple times and includes the instructions to be parallelized through pipelining; its length is the $II$.
A high quality mapping is associated with a low $II$, with the focus of mapping methods being to find the lowest possible $II$ for a given DFG. An example of a legal mapping for the DFG in our running example is shown in Fig.~\ref{fig:example_b}.

\subsection{Mapping Problem}

We tackle the mapping problem by dividing it into two distinct phases: time and space. First, we focus on finding a valid time solution, meaning that we aid to find a schedule that correctly resolves all the data dependencies in the DFG. Once the time solution is found, we move to the spatial phase, where all the placement of operations and routing of data is done accordingly to the CGRA size and topology.
Fig.~\ref{fig:example_c} illustrates valid and invalid solutions in time and space. On the figure's right, we have an invalid schedule for nodes 2 and 8 from the running example, with both nodes being scheduled at the same time step despite a dependency between them. An invalid spatial mapping is also shown, in which node 4 is placed on PE0 and node 7 on PE3. Due to the topology of the CGRA, the routing between PE0 and PE3 is not possible, causing the loop-carried dependency to be violated.


\section{Methodology} \label{sec:methodology}

Decoupling space and time allows the search for a CGRA mapping to be done in two phases. Thus, it is possible to perform an independent search in one dimension first and then find an associated solution in the other one. Our approach first finds a solution in time, via a modified version of SAT-MapIt's formulation~\cite{Tirelli2023}, and then finds a solution in space, using monomorphism~\cite{bonnici2013subgraph,carletti2017challenging}. We start by defining the compilation components, in Section~\ref{sec:methodology_definitions}, and then describe how a solution in time can be found, in Section~\ref{sec:methodology_time}, and how this solution can be extended to a solution in space, in Section~\ref{sec:methodology_space}, lastly, in Section~\ref{sec:methodology_proof}, we provide a proof that a time solution under our constraints always implies the existence of a space solution.

\subsection{Formal Definitions} \label{sec:methodology_definitions}

We define a DFG as an undirected graph $\mathcal{G} = (\mathcal{V}_\mathcal{G}, \mathcal{E}_\mathcal{G}, l_\mathcal{G})$, where $\mathcal{V}_\mathcal{G}$ is the set of vertices, $\mathcal{E}_\mathcal{G} \subseteq \{\{u,v\} : u,v \in \mathcal{V}_\mathcal{G}\}$ is the set of edges, and $l_\mathcal{G} : \mathcal{V}_\mathcal{G} \rightarrow \mathcal{L}$ is a labeling function with $\mathcal{L} = \{0, ...,II-1\}$ and $II \in \mathbb{N}^+$. This is the structure that we wish to compile to the CGRA.

\begin{figure}[t]
    \centering
    \includegraphics[width=0.3\textwidth]{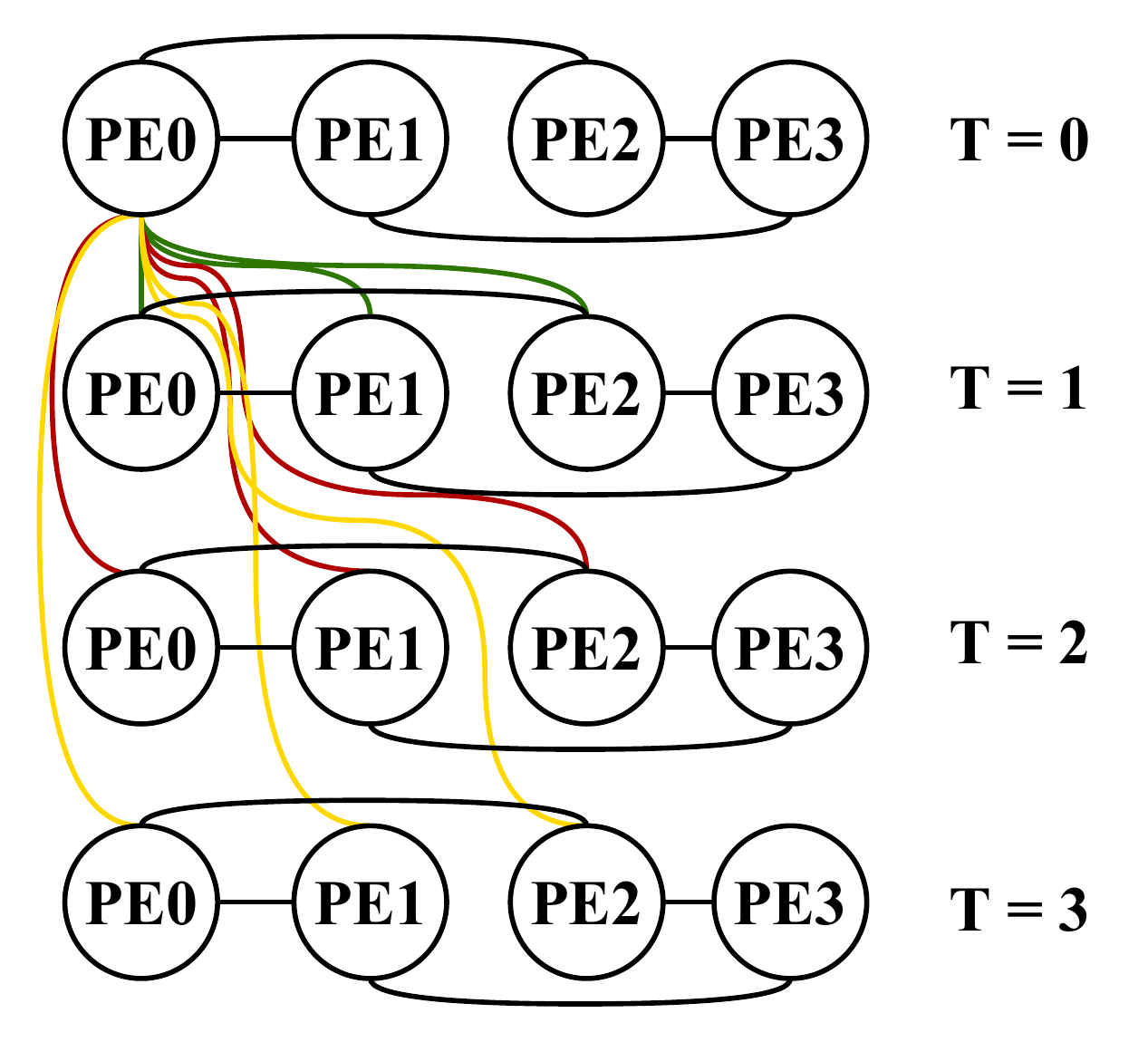}
    \caption{MRRG for a $2 \times 2$ CGRA and $II = 4$. Black edges represent CGRA adjacencies, while green, red, and yellow edges represent time adjacencies from PE0 at $T = 0$. Time adjacencies of the other PEs, as well as the self-loops inherent to every PE, are omitted for clarity.}
    \label{fig:mrrg}
\end{figure}

To represent the allocation of instructions to the CGRA through time, we need to define an MRRG, whose structure consists of $II$ stacked copies of the CGRA architecture linked via all valid routes allowed by the CGRA's topology. Fig.~\ref{fig:mrrg} shows an example of MRRG for a $2\times 2$ CGRA with $II = 4$.
Each one of the $II$ architecture copies is defined as $\mathcal{M}^i = (\mathcal{V}_\mathcal{M^\textnormal{i}}, \mathcal{E}_\mathcal{M^\textnormal{i}})$, where $\mathcal{V}_\mathcal{M^\textnormal{i}}$ is the set of vertices representing the CGRA's PEs at time step $i$ and $\mathcal{E}_\mathcal{M^\textnormal{i}} \subseteq \{\{u,v\} : u,v \in \mathcal{V}^i_\mathcal{M}\}$ is the set of edges representing the interconnections among the PEs of the CGRA. In Fig.~\ref{fig:mrrg}, the vertices at $T = 0$ would be in the set $\mathcal{M}^0$ and all the black edges at that time step would be in $\mathcal{E}_{\mathcal{M}^0}$.

We define the MRRG as $\mathcal{M} = (\mathcal{V}_\mathcal{M}, \mathcal{E}_\mathcal{M}, l_\mathcal{M})$, where $\mathcal{V}_\mathcal{M}$ has the CGRA's PEs for all available time steps and $\mathcal{E}_\mathcal{M}$ has all interconnections between PEs at every time step, as follows:
\begin{align*}
    \mathcal{V}_\mathcal{M} &= \bigcup_{i \in \mathcal{L}} \mathcal{V}_\mathcal{M^\textnormal{i}}
    &
    \mathcal{E}_\mathcal{M} &= 
        \bigcup_{i \in \mathcal{L}} \mathcal{E}_\mathcal{M^\textnormal{i}}
        \cup
        \bigcup_{i,j \in \mathcal{L}}
            \curlyBegin e : \land
                \blockBegin
                e \in \mathcal{E}_\mathcal{M^\textnormal{i},M^\textnormal{j}} \\
                i = j - 1
                \blockEnd
            \curlyEnd
\end{align*}
with $\mathcal{E}_\mathcal{M^\textnormal{i},M^\textnormal{j}} \subseteq \{\{u,v\} : u \in \mathcal{V}_\mathcal{M^\textnormal{i}}, v \in \mathcal{V}_\mathcal{M^\textnormal{j}}\}$.
The set $\mathcal{E}_\mathcal{M^\textnormal{0},M^\textnormal{1}}$ is partially depicted in green in Fig.~\ref{fig:mrrg} for the vertex associate to PE0.
The function $l_\mathcal{M} : \mathcal{V}_\mathcal{M} \rightarrow \mathcal{L}$ assigns a fixed label to every vertex, reflecting the fact that $\mathcal{V}_\mathcal{M^\textnormal{i}}$ represents the CGRA architecture at time step $i$. In Fig.~\ref{fig:mrrg} all vertices at $T = 0$ have label 0, while the ones at $T = 1$ have label 1, and so on.
Concretely, we have that $\forall i \in \mathcal{L} \centerdot \forall v \in \mathcal{V}_\mathcal{M^\textnormal{i}} \centerdot l_\mathcal{M}(v) = i$.

A property of interest for compilation is the connectivity degree of vertices in the MRRG. For every vertex $v \in \mathcal{V}_\mathcal{M^\textnormal{i}}$ and time step $i \in \mathcal{L}$, we define the connectivity degree of $v$ as $\mathcal{D}^v_\mathcal{M^\textnormal{i}} = |\{\{u,v\} \in \mathcal{E}_\mathcal{M^\textnormal{i}}\}|$. Since all the vertices of $\mathcal{M}$ have the same degree, we have $\forall i \in \mathcal{L} \centerdot \forall v \in \mathcal{V}_\mathcal{M^\textnormal{i}} \centerdot \mathcal{D}_\mathcal{M} = \mathcal{D}_\mathcal{M^\textnormal{i}} = \mathcal{D}^v_\mathcal{M^\textnormal{i}}$. An MRRG with connectivity degree 3 can be seen in Fig.~\ref{fig:mrrg}; self-loops are in the MRRG but are omitted for clarity.

\begin{figure}[t]
    \centering
    \includegraphics[width=0.3\textwidth]{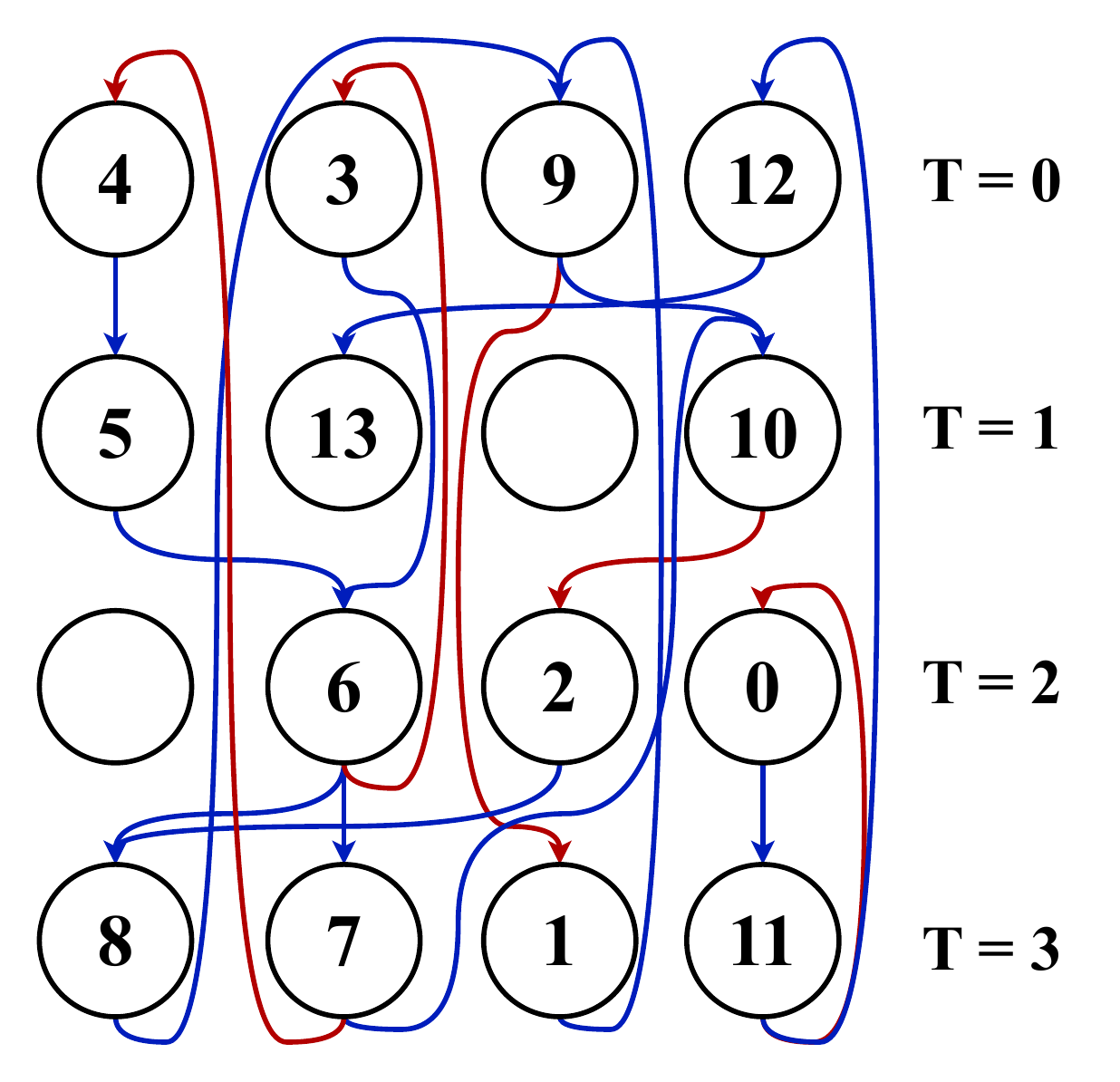}
    \caption{Monomorphism between the DFG shown in Fig.~\ref{fig:example} and the MRRG shown in Fig.~\ref{fig:mrrg}; data dependencies are in blue and loop-carried dependencies routed in the MRRG are in red.}
    \label{fig:mono}
\end{figure}

The final element we need to define is monomorphism, which allows us to find a solution in space given a solution in time. A monomorphism from a graph $\mathcal{G}$ to a graph $\mathcal{M}$ is a function $f : \mathcal{V}_\mathcal{G} \rightarrow \mathcal{V}_\mathcal{M}$ that respects three properties:
\begin{subequations}
    \begin{empheq}{align}
        f &\text{ is injective}
        \tag{mono1} \label{eq:mono_cond1} \\
        \forall v \in \mathcal{V}_\mathcal{G} &\centerdot l_\mathcal{G}(v) = l_\mathcal{M}(f(v))
        \tag{mono2} \label{eq:mono_cond2} \\
        \forall \{u,v\} \in \mathcal{E}_\mathcal{G} &\centerdot \{f(u),f(v)\} \in \mathcal{E}_\mathcal{M}
        \tag{mono3} \label{eq:mono_cond3}
    \end{empheq}
\end{subequations}
Property~\ref{eq:mono_cond1} ensures that one vertex of the MRRG is the target of at most one vertex of the DFG, since one PE can execute only one operation at any given time step. Property~\ref{eq:mono_cond2} ensures that every vertex is executed at the correct time step. Property~\ref{eq:mono_cond3} ensures that an MRRG edge satisfies the data dependencies. An example can be seen in Fig.~\ref{fig:mono}.

\subsection{Time Solution} \label{sec:methodology_time}

In our formulation, we define the DFG as an undirected graph. This is possible because we initially find a schedule with the directed version of the DFG. Once a schedule is found, each vertex is labeled with its respective timing information and the directionality of the edges becomes redundant and is removed.

To find a suitable time schedule, we start by creating the as soon as possible (ASAP) and as late as possible (ALAP) schedules for the input DFG, to derive the range of possible scheduling time steps of every vertex. This is used in the Mobility Schedule (MobS), which expresses the mobility of each vertex. Tab.~\ref{tab:scheds} shows these schedules for the DFG in Fig.~\ref{fig:example}.

\begin{table}[b]
    \centering
    \caption{ASAP, ALAP, and MobS for the DFG in Fig.~\ref{fig:example}}
    \label{tab:scheds}
    \begin{tabular}{clll}
        \toprule
         & \multicolumn{3}{c}{\textbf{Nodes}} \\
         \cmidrule(l){2-4}
         \textbf{Time} & \multicolumn{1}{c}{\textbf{ASAP}} & \multicolumn{1}{c}{\textbf{ALAP}} & \multicolumn{1}{c}{\textbf{MobS}} \\
         \cmidrule(l){1-1} \cmidrule(l){2-4}
         0 & 0 1 2 3 4 & 4      & 0 1 2 3 4      \\ 
         1 & 5 11      & 3 5    & 0 1 2 3 5 11   \\
         2 & 6 12      & 0 2 6  & 0 1 2 6 11 12  \\
         3 & 7 8 13    & 1 8 11 & 1 7 8 11 12 13 \\
         4 & 9         & 7 9 12 & 7 9 12 13      \\ 
         5 & 10        & 10 13  & 10 13          \\ 
         \bottomrule
    \end{tabular}
\end{table}

Similar to the approach of Tirelli et al.~\cite{Tirelli2023}, the MobS is the base structure used, together with the $II$, to create the Kernel Mobility Schedule (KMS) needed to formulate the scheduling problem. The KMS is the \emph{superset of all possible schedules} for a given $II$, and is the result of iteratively folding the MobS by an amount equal to $II$.
After each folding happens, every vertex copied into the KMS is assigned a label that refers to the iteration number. The number of loop iterations interleaved and executed at the same time is $\lceil \frac{MobS_{length}}{II} \rceil$. In our example, we have $\lceil 6/4 \rceil = 2$, thus every execution of the kernel computes data from two different loop iterations. Tab.~\ref{tab:kmsgen} shows the KMS for the DFG in Fig.~\ref{fig:example}.

The search starts from the $mII$, defined in \cite{rau1996iterative} as follows:
\begin{align*}
    mII = max(ResII, RecII)
\end{align*}
\begin{align*}
    ResII &= \ceilBegin \frac{|\mathcal{V}_\mathcal{G}|}{|\mathcal{V}_\mathcal{M^\textnormal{i}}|} \ceilEnd
    &
    RecII &= \max_{l \in DFG} \ceilBegin \frac{length(l)}{distance(l)} \ceilEnd
\end{align*}
where $ResII$ represents the minimum number of resources required by the architecture to execute all the instructions in the DFG, and $RecII$ represents the longest cycle length in the DFG. In our example, we have $ResII = \lceil \frac{14}{2\cdot 2} \rceil = 4$ and $RecII = 4$. Consequentially, $mII = max(4,4) = 4$ is our starting point, meaning that no solution below $II = 4$ is valid.

\begin{table}[b]
    \centering
    \caption{KMS for the MobS in Tab.~\ref{tab:scheds} and an $II$ of 4.}
    \label{tab:kmsgen}
    \begin{tabular}{cll}
        \toprule
        \textbf{Time} & \multicolumn{2}{c}{\textbf{Nodes}} \\
        \cmidrule(l){1-1} \cmidrule(l){2-3}
         0 & $ 0_0 1_0 2_0 6_0 11_0 12_0$ &                            \\
         1 & $1_0 7_0 8_0 11_0 12_0 13_0$ &                            \\
         2 & $7_0 9_0 12_0 13_0$          & $0_1 1_1 2_1 3_1 4_1$      \\
         3 & $10_0 13_0$                  & $0_1 1_1 2_1 3_1 5_1 11_1$ \\
         \bottomrule
    \end{tabular}
\end{table}

Inspired by the idea used in~\cite{Tirelli2023} to express the solution space of the mapping problem, we design our own form of expressing the solution space which focus solely on the time dimension.
We use the KMS to describe the solution space of the time dimension as a Satisfiability Modulo Theories (SMT) formula and include additional constraints, not present in~\cite{Tirelli2023}, to ensure that the schedule obtained in our search leads to a monomorphism of the DFG in the MRRG. The constraints of our formulation are divided in three sets: modulo scheduling, capacity, and connectivity; the latter two are our additions.

\subsubsection{Modulo scheduling constraints} \label{sec:methodology_time_scheduling}

The labeling of the vertices corresponds to their scheduling. With the following constraints, we ensure that the correct order of execution of all the DFG vertices is respected. Data dependencies and loop-carried dependencies are respectively encoded as follows:
\begin{align*}
    \left\{
        \begin{array}{@{}ll@{}}
            t_d \leq t_s \text{ if } it_s - it_d = 1
            \\
            t_d > t_s \text{ if } it_s = it_d
        \end{array}
    \right.
    & & & &
    \left\{
        \begin{array}{@{}ll@{}}
            t_d \leq t_s \text{ if } it_s = it_d
            \\
            t_d > t_s \text{ if } it_s - it_d = 1
        \end{array}
    \right.
\end{align*}
where $t_d$ and $t_s$ are the times at which the destination and source vertices can be scheduled, and $it_s$ and $it_d$ refer to the subscripts in the KMS that indicate how many foldings have been done on the MobS. In the KMS there are many possibilities, not all of them are valid. With our SMT formulation, we can explore the scheduling space to find all valid solutions.

\subsubsection{Capacity constraints} \label{sec:methodology_time_capacity}

These constraints ensure that, once a valid schedule is found in the time dimension, it can be effectively mapped onto the physical resources of the CGRA. Concretely, we need to ensure that the capacity of the CGRA is not exceeded at any time step. Thus, we require that $\forall i \in \mathcal{L} \centerdot C_i \leq |\mathcal{V}_\mathcal{M^\textnormal{i}}|$, where $C_i = |\{u \in \mathcal{V}_\mathcal{G} : l_\mathcal{G}(u) = i\}|$.

\subsubsection{Connectivity constraints} \label{sec:methodology_time_connectivity}

We also need to ensure that the connectivity degree of the PEs is never exceeded. Let $S^i_v$ be the set of neighbors of $v \in \mathcal{V}_\mathcal{G}$ at time step $i \in \mathcal{L}$, defined as $S^i_v = \{\{u,v\} \in \mathcal{E}_\mathcal{G} : l_\mathcal{G}(u) = i\}$. Having $\mathcal{D}_\mathcal{M}$ as the CGRA connectivity degree, e.g., $\mathcal{D}_\mathcal{M} = 3$ in a $2\times2$ architecture and $\mathcal{D}_\mathcal{M} = 5$ in $3\times3$ and larger architectures, we must require that $\forall v \in \mathcal{V}_\mathcal{G} \centerdot \forall i \in \mathcal{L} \centerdot |S^i_v| \leq \mathcal{D}_\mathcal{M}$. This allows for the correct routing of resources for every PE, since the number of successors of a DFG vertex scheduled on the same time step will be at most the number of neighbors of a PE.

\subsection{Space Solution} \label{sec:methodology_space}

Many techniques found in the literature can be reused to navigate the spatial dimension. We, however, follow a novel approach based on monomorphism extraction. Algorithms for such extraction~\cite{carletti2017challenging, bonnici2013subgraph} have demonstrated their ability to handle large graphs with low computational overhead, which is corroborated by our experimental results.

\subsection{Proof of Monomorphism Existence} \label{sec:methodology_proof}

We want to prove that, for all $\mathcal{G}$ and $\mathcal{M}$ satisfying the above constraints, there exists a monomorphism $f$ from $\mathcal{G}$ to $\mathcal{M}$. Concretely, we need to show that there exists a function that respects properties \ref{eq:mono_cond1}, \ref{eq:mono_cond2}, and \ref{eq:mono_cond3}.

\subsubsection{Function $f$ is injective}

To establish that there exists an injective interpretation for the function $f : \mathcal{V}_\mathcal{G} \rightarrow \mathcal{V}_\mathcal{M}$ we must show that its codomain it at least as large as its domain. This is guaranteed by two following disequalities:
\begin{align*}
    |\mathcal{V}_\mathcal{G}| = \sum_{i \in \mathcal{L}} C_i &\leq \sum_{i \in \mathcal{L}} |\mathcal{V}_\mathcal{M^\textnormal{i}}| = |\mathcal{V}_\mathcal{M}|
    & & &
    \forall i \in \mathcal{L} \centerdot C_i &\leq |\mathcal{V}_\mathcal{M^\textnormal{i}}|
\end{align*}

\subsubsection{Function $f$ maps vertices}

To establish that a vertex mapping is possible we must show that the number of vertices labeled with a specific $i \in \mathcal{L}$, i.e., $C_i$, is at most the capacity of the CGRA, i.e., $|\mathcal{V}_\mathcal{M^\textnormal{i}}|$. This is guaranteed by the capacity constraints described in Section~\ref{sec:methodology_time_capacity}.

\subsubsection{Function $f$ maps edges}

To establish that an edge mapping is possible we must show that the cardinality of the set of neighbors of every vertex $v \in \mathcal{V}_\mathcal{G}$ labeled with a specific $i \in \mathcal{L}$, i.e., $|S^i_v|$, is at most the connectivity degree of the CGRA, i.e., $\mathcal{D}_\mathcal{M}$. This is guaranteed by the connectivity constraints described in Section~\ref{sec:methodology_time_connectivity}.


\section{Experiments}

\begin{figure}[b]
    \centering
    \includegraphics[width=\linewidth]{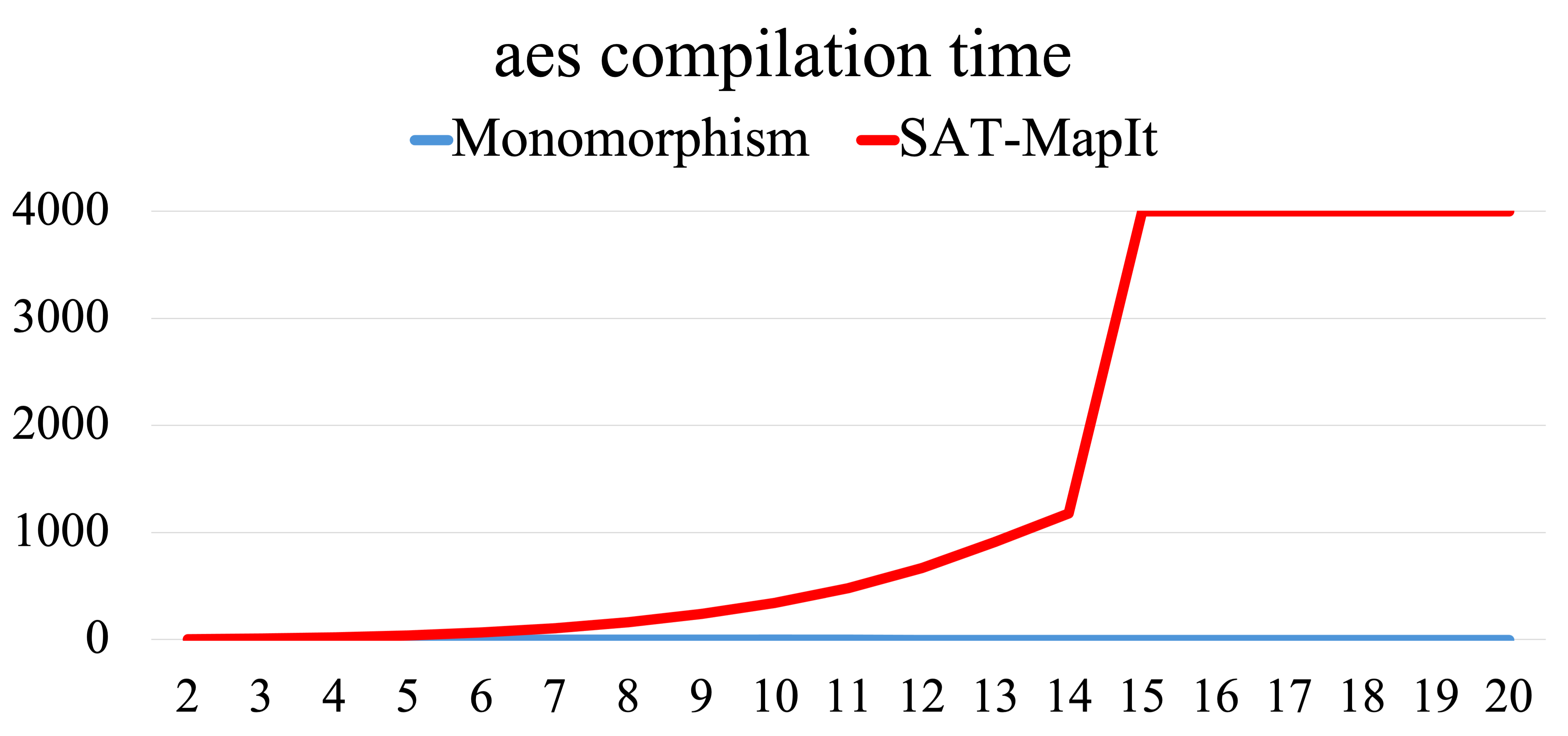}
    \caption{Compilation time (y-axis), in seconds, in relation to CGRA sizes (x-axis) for the aes benchmark.}
    \label{fig:aesmonosat}
\end{figure}

We evaluated our methodology on four different CGRA configurations: $2 \times 2$, $5 \times 5$, $10 \times 10$, and $20 \times 20$. Each PE register file in the CGRA could be accessed by neighboring PEs. We choose all the innermost loops from the MiBench~\cite{mibench} and Rodina~\cite{rodinia} benchmarks suites, excluding those with function calls or conditional statements inside the loop body, resulting in a total of 17 benchmarks. We compared our approach against that of~\cite{Tirelli2023}, since it is compatible with our target architecture and was shown to consistently yield better results than other comparable state-of-the-art methodologies. The metrics chosen for our evaluation were $II$ value and compilation time. We used the Z3 solver~\cite{moura2008z3} to solve the SMT formulas generated. All experiments were performed on a Linux Machine with a 3.30~GHz Intel Core i9 CPU and 256~GB of RAM.

\subsubsection{Iteration Interval comparison}

Tab.~\ref{tab:benchres} shows the $II$ values obtained in our experiments, with solutions being found by our methodology in 62 out of 68 cases.
Our approach achieves the same $II$ as~\cite{Tirelli2023} in 57 cases. In 5 cases, we find a solution, but the tool from~\cite{Tirelli2023} times out, which demonstrates our ability to obtain quality  mappings. In only one case, out of all our 68 experiments, our space search reaches a timeout but the compared tool does not.

\begin{table*}[t]
    \centering
    \setlength{\tabcolsep}{1mm}
    \begin{tabular}{lrrrrrrrrrrrrrrr}
    \toprule
    & \multicolumn{1}{c}{}
    & \multicolumn{7}{c}{$2 \times 2$ CGRA}
    & \multicolumn{7}{c}{$5 \times 5$ CGRA}
    \\
    \cmidrule(l){3-9} \cmidrule(l){10-16}
    & \multicolumn{1}{c}{}
    & \multicolumn{3}{c}{\textbf{Compilation Time}}
    & \multicolumn{1}{c}{\multirow{3}{*}{\textbf{$\Delta$T}}}
    & \multicolumn{1}{c}{\multirow{3}{*}{\textbf{CTR}}}
    & \multicolumn{1}{c}{\multirow{3}{*}{\textit{\textbf{II}}}}
    & \multicolumn{1}{c}{\multirow{3}{*}{\textit{\textbf{mII}}}}
    & \multicolumn{3}{c}{\textbf{Compilation Time}}
    & \multicolumn{1}{c}{\multirow{3}{*}{\textbf{$\Delta$T}}}
    & \multicolumn{1}{c}{\multirow{3}{*}{\textbf{CTR}}}
    & \multicolumn{1}{c}{\multirow{3}{*}{\textit{\textbf{II}}}}
    & \multicolumn{1}{c}{\multirow{3}{*}{\textit{\textbf{mII}}}}
    \\
    \cmidrule(l){3-5} \cmidrule(l){10-12}
    \multicolumn{1}{c}{\multirow{2}{*}{\textbf{Benchmark}}}
    & \multicolumn{1}{c}{\multirow{2}{*}{\textbf{\begin{tabular}[c]{@{}c@{}}DFG \\ Nodes \end{tabular}}}}
    & \multicolumn{2}{c}{\textbf{Monomorphism}}
    & \multicolumn{1}{c}{\multirow{2}{*}{\textbf{SAT-MapIt}}}
    & \multicolumn{1}{c}{}
    & \multicolumn{1}{c}{}
    & \multicolumn{1}{c}{}
    & \multicolumn{1}{c}{}
    & \multicolumn{2}{c}{\textbf{Monomorphism}}
    & \multicolumn{1}{c}{\multirow{2}{*}{\textbf{SAT-MapIt}}}
    & \multicolumn{1}{c}{}
    & \multicolumn{1}{c}{}
    & \multicolumn{1}{c}{}
    & \multicolumn{1}{c}{}
    \\
    \cmidrule(l){3-4} \cmidrule(l){10-11}
    & \multicolumn{1}{c}{}
    & \multicolumn{1}{c}{\textbf{Time}}
    & \multicolumn{1}{c}{\textbf{Space}}
    & \multicolumn{1}{c}{}
    & \multicolumn{1}{c}{}
    & \multicolumn{1}{c}{}
    & \multicolumn{1}{c}{}
    & \multicolumn{1}{c}{}
    & \multicolumn{1}{c}{\textbf{Time}}
    & \multicolumn{1}{c}{\textbf{Space}}
    & \multicolumn{1}{c}{}
    & \multicolumn{1}{c}{}
    & \multicolumn{1}{c}{}
    & \multicolumn{1}{c}{}
    & \multicolumn{1}{c}{}
    \\
    \cmidrule(l){1-1} \cmidrule(l){2-2} \cmidrule(l){3-9} \cmidrule(l){10-16}
    \multicolumn{1}{l}{aes}
        & 23
        & \multicolumn{1}{r}{0.40}
        & \multicolumn{1}{r}{0.02}
        & \multicolumn{1}{r}{2.57}
        & \multicolumn{1}{r}{-2.15}
        & 6.12
        & \multicolumn{1}{r}{16}
        & 14
        & \multicolumn{1}{r}{0.47}
        & \multicolumn{1}{r}{0.04}
        & \multicolumn{1}{r}{39.07}
        & \multicolumn{1}{r}{-38.56}
        & 76.16
        & \multicolumn{1}{r}{16}
        & 14
        \\
    \multicolumn{1}{l}{backprop}
        & 34
        & \multicolumn{1}{r}{0.44}
        & \multicolumn{1}{r}{0.03}
        & \multicolumn{1}{r}{110.01}
        & \multicolumn{1}{r}{-109.54}
        & 233.07
        & \multicolumn{1}{r}{10}
        & 9
        & \multicolumn{1}{r}{0.12}
        & \multicolumn{1}{r}{0.29}
        & \multicolumn{1}{r}{9.98}
        & \multicolumn{1}{r}{-9.56}
        & 23.82
        & \multicolumn{1}{r}{5}
        & 5
        \\
    \multicolumn{1}{l}{basicmath}
        & 21
        & \multicolumn{1}{r}{0.32}
        & \multicolumn{1}{r}{0.11}
        & \multicolumn{1}{r}{0.42}
        & \multicolumn{1}{r}{0.01}
        & 0.98
        & \multicolumn{1}{r}{7}
        & 7
        & \multicolumn{1}{r}{0.13}
        & \multicolumn{1}{r}{0.31}
        & \multicolumn{1}{r}{7.82}
        & \multicolumn{1}{r}{-7.38}
        & 17.77
        & \multicolumn{1}{r}{7}
        & 7
        \\
    \multicolumn{1}{l}{bitcount}
        & 7
        & \multicolumn{1}{r}{0.038}
        & \multicolumn{1}{r}{$\sim$0.01}
        & \multicolumn{1}{r}{0.06}
        & \multicolumn{1}{r}{-0.02}
        & 1.73
        & \multicolumn{1}{r}{3}
        & 3
        & \multicolumn{1}{r}{0.39}
        & \multicolumn{1}{r}{$\sim$0.01}
        & \multicolumn{1}{r}{1.15}
        & \multicolumn{1}{r}{-0.76}
        & 2.95
        & \multicolumn{1}{r}{3}
        & 3
        \\
    \multicolumn{1}{l}{cfd}
        & 51
        & \multicolumn{1}{r}{TO}
        & \multicolumn{1}{r}{-}
        & \multicolumn{1}{r}{TO}
        & \multicolumn{1}{r}{-}
        & -
        & \multicolumn{1}{r}{-}
        & 13
        & \multicolumn{1}{r}{0.07}
        & \multicolumn{1}{r}{TO}
        & \multicolumn{1}{r}{23.59}
        & \multicolumn{1}{r}{-}
        & -
        & \multicolumn{1}{r}{3}
        & 3
        \\
    \multicolumn{1}{l}{crc32}
        & 24
        & \multicolumn{1}{r}{0.20}
        & \multicolumn{1}{r}{$\sim$0.01}
        & \multicolumn{1}{r}{3.85}
        & \multicolumn{1}{r}{-3.64}
        & 18.34
        & \multicolumn{1}{r}{11}
        & 8
        & \multicolumn{1}{r}{0.30}
        & \multicolumn{1}{r}{$\sim$0.01}
        & \multicolumn{1}{r}{75.75}
        & \multicolumn{1}{r}{-75.45}
        & 250.00
        & \multicolumn{1}{r}{11}
        & 8
        \\
    \multicolumn{1}{l}{fft}
        & 20
        & \multicolumn{1}{r}{0.09}
        & \multicolumn{1}{r}{$\sim$0.01}
        & \multicolumn{1}{r}{0.46}
        & \multicolumn{1}{r}{-0.37}
        & 5.05
        & \multicolumn{1}{r}{7}
        & 7
        & \multicolumn{1}{r}{0.14}
        & \multicolumn{1}{r}{$\sim$0.01}
        & \multicolumn{1}{r}{8.22}
        & \multicolumn{1}{r}{-8.08}
        & 57.89
        & \multicolumn{1}{r}{7}
        & 7
        \\
    \multicolumn{1}{l}{gsm}
        & 24
        & \multicolumn{1}{r}{0.06}
        & \multicolumn{1}{r}{$\sim$0.01}
        & \multicolumn{1}{r}{0.43}
        & \multicolumn{1}{r}{-0.36}
        & 6.30
        & \multicolumn{1}{r}{6}
        & 6
        & \multicolumn{1}{r}{0.11}
        & \multicolumn{1}{r}{$\sim$0.01}
        & \multicolumn{1}{r}{15.49}
        & \multicolumn{1}{r}{-15.36}
        & 122.94
        & \multicolumn{1}{r}{5}
        & 4
        \\
    \multicolumn{1}{l}{heartwall}
        & 35
        & \multicolumn{1}{r}{0.14}
        & \multicolumn{1}{r}{$\sim$0.01}
        & \multicolumn{1}{r}{1.31}
        & \multicolumn{1}{r}{-1.17}
        & 9.21
        & \multicolumn{1}{r}{9}
        & 9
        & \multicolumn{1}{r}{0.16}
        & \multicolumn{1}{r}{$\sim$0.01}
        & \multicolumn{1}{r}{45.18}
        & \multicolumn{1}{r}{-45.01}
        & 272.17
        & \multicolumn{1}{r}{3}
        & 3
        \\
    \multicolumn{1}{l}{hotspot3D}
        & 57
        & \multicolumn{1}{r}{1.13}
        & \multicolumn{1}{r}{0.09}
        & \multicolumn{1}{r}{223.51}
        & \multicolumn{1}{r}{-222.29}
        & 182.46
        & \multicolumn{1}{r}{17}
        & 15
        & \multicolumn{1}{r}{0.54}
        & \multicolumn{1}{r}{0.01}
        & \multicolumn{1}{r}{209.87}
        & \multicolumn{1}{r}{-209.32}
        & 378.14
        & \multicolumn{1}{r}{6}
        & 3
        \\
    \multicolumn{1}{l}{lud}
        & 26
        & \multicolumn{1}{r}{0.07}
        & \multicolumn{1}{r}{$\sim$0.01}
        & \multicolumn{1}{r}{0.45}
        & \multicolumn{1}{r}{-0.37}
        & 5.54
        & \multicolumn{1}{r}{7}
        & 7
        & \multicolumn{1}{r}{0.07}
        & \multicolumn{1}{r}{$\sim$0.01}
        & \multicolumn{1}{r}{7.95}
        & \multicolumn{1}{r}{-7.88}
        & 107.72
        & \multicolumn{1}{r}{3}
        & 3
        \\
    \multicolumn{1}{l}{nw}
        & 33
        & \multicolumn{1}{r}{0.18}
        & \multicolumn{1}{r}{$\sim$0.01}
        & \multicolumn{1}{r}{2.48}
        & \multicolumn{1}{r}{-2.29}
        & 13.03
        & \multicolumn{1}{r}{9}
        & 9
        & \multicolumn{1}{r}{0.05}
        & \multicolumn{1}{r}{1.16}
        & \multicolumn{1}{r}{5.39}
        & \multicolumn{1}{r}{-4.17}
        & 4.43
        & \multicolumn{1}{r}{2}
        & 2
        \\
    \multicolumn{1}{l}{particlefilter}
        & 38
        & \multicolumn{1}{r}{0.12}
        & \multicolumn{1}{r}{$\sim$0.01}
        & \multicolumn{1}{r}{1.67}
        & \multicolumn{1}{r}{-1.55}
        & 13.47
        & \multicolumn{1}{r}{10}
        & 10
        & \multicolumn{1}{r}{0.34}
        & \multicolumn{1}{r}{$\sim$0.01}
        & \multicolumn{1}{r}{28.08}
        & \multicolumn{1}{r}{-27.73}
        & 81.16
        & \multicolumn{1}{r}{9}
        & 9
        \\
    \multicolumn{1}{l}{sha1}
        & 21
        & \multicolumn{1}{r}{0.05}
        & \multicolumn{1}{r}{0.43}
        & \multicolumn{1}{r}{0.27}
        & \multicolumn{1}{r}{0.21}
        & 0.56
        & \multicolumn{1}{r}{6}
        & 6
        & \multicolumn{1}{r}{0.11}
        & \multicolumn{1}{r}{0.09}
        & \multicolumn{1}{r}{15.44}
        & \multicolumn{1}{r}{-15.24}
        & 77.20
        & \multicolumn{1}{r}{4}
        & 2
        \\
    \multicolumn{1}{l}{sha2}
        & 25
        & \multicolumn{1}{r}{0.07}
        & \multicolumn{1}{r}{$\sim$0.01}
        & \multicolumn{1}{r}{0.60}
        & \multicolumn{1}{r}{-0.52}
        & 7.29
        & \multicolumn{1}{r}{7}
        & 6
        & \multicolumn{1}{r}{0.16}
        & \multicolumn{1}{r}{4.07}
        & \multicolumn{1}{r}{9.22}
        & \multicolumn{1}{r}{-4.99}
        & 2.18
        & \multicolumn{1}{r}{7}
        & 7
        \\
    \multicolumn{1}{l}{stringsearch}
        & 28
        & \multicolumn{1}{r}{0.10}
        & \multicolumn{1}{r}{$\sim$0.01}
        & \multicolumn{1}{r}{1.04}
        & \multicolumn{1}{r}{-0.94}
        & 9.90
        & \multicolumn{1}{r}{7}
        & 7
        & \multicolumn{1}{r}{0.10}
        & \multicolumn{1}{r}{1.09}
        & \multicolumn{1}{r}{17.01}
        & \multicolumn{1}{r}{-15.82}
        & 14.29
        & \multicolumn{1}{r}{3}
        & 3
        \\
    \multicolumn{1}{l}{susan}
        & 21
        & \multicolumn{1}{r}{0.09}
        & \multicolumn{1}{r}{$\sim$0.01}
        & \multicolumn{1}{r}{0.97}
        & \multicolumn{1}{r}{-0.88}
        & 10.34
        & \multicolumn{1}{r}{6}
        & 6
        & \multicolumn{1}{r}{0.08}
        & \multicolumn{1}{r}{$\sim$0.01}
        & \multicolumn{1}{r}{15.94}
        & \multicolumn{1}{r}{-15.85}
        & 171.40
        & \multicolumn{1}{r}{2}
        & 2
        \\
    \cmidrule(l){1-1} \cmidrule(l){2-2} \cmidrule(l){3-9} \cmidrule(l){10-16}
    \multicolumn{1}{l}{Average}
        & -
        & \multicolumn{1}{r}{0.22}
        & \multicolumn{1}{r}{0.042}
        & \multicolumn{1}{r}{21.88}
        & \multicolumn{1}{r}{\textbf{-21.61}}
        & \textbf{30.85} 
        & \multicolumn{1}{r}{-}
        & -
        & \multicolumn{1}{r}{0.20}
        & \multicolumn{1}{r}{0.44}
        & \multicolumn{1}{r}{31.97}
        & \multicolumn{1}{r}{\textbf{-31.32}}
        & \textbf{103.76} 
        & \multicolumn{1}{r}{-}
        & -
        \\
        \toprule
    & \multicolumn{1}{c}{}
    & \multicolumn{7}{c}{$10 \times 10$ CGRA}
    & \multicolumn{7}{c}{$20 \times 20$ CGRA}
    \\
    \cmidrule(l){3-9} \cmidrule(l){10-16}
    & \multicolumn{1}{c}{}
    & \multicolumn{3}{c}{\textbf{Compilation Time}}
    & \multicolumn{1}{c}{\multirow{3}{*}{\textbf{$\Delta$T}}}
    & \multicolumn{1}{c}{\multirow{3}{*}{\textbf{CTR}}}
    & \multicolumn{1}{c}{\multirow{3}{*}{\textit{\textbf{II}}}}
    & \multicolumn{1}{c}{\multirow{3}{*}{\textit{\textbf{mII}}}}
    & \multicolumn{3}{c}{\textbf{Compilation Time}}
    & \multicolumn{1}{c}{\multirow{3}{*}{\textbf{$\Delta$T}}}
    & \multicolumn{1}{c}{\multirow{3}{*}{\textbf{CTR}}}
    & \multicolumn{1}{c}{\multirow{3}{*}{\textit{\textbf{II}}}}
    & \multicolumn{1}{c}{\multirow{3}{*}{\textit{\textbf{mII}}}}
    \\
    \cmidrule(l){3-5} \cmidrule(l){10-12}
    \multicolumn{1}{c}{\multirow{2}{*}{\textbf{Benchmark}}}
    & \multicolumn{1}{c}{\multirow{2}{*}{\textbf{\begin{tabular}[c]{@{}c@{}}DFG \\ Nodes \end{tabular}}}}
    & \multicolumn{2}{c}{\textbf{Monomorphism}}
    & \multicolumn{1}{c}{\multirow{2}{*}{\textbf{SAT-MapIt}}}
    & \multicolumn{1}{c}{}
    & \multicolumn{1}{c}{}
    & \multicolumn{1}{c}{}
    & \multicolumn{1}{c}{}
    & \multicolumn{2}{c}{\textbf{Monomorphism}}
    & \multicolumn{1}{c}{\multirow{2}{*}{\textbf{SAT-MapIt}}}
    & \multicolumn{1}{c}{}
    & \multicolumn{1}{c}{}
    & \multicolumn{1}{c}{}
    & \multicolumn{1}{c}{}
    \\
    \cmidrule(l){3-4} \cmidrule(l){10-11}
    & \multicolumn{1}{c}{}
    & \multicolumn{1}{c}{\textbf{Time}}
    & \multicolumn{1}{c}{\textbf{Space}}
    & \multicolumn{1}{c}{}
    & \multicolumn{1}{c}{}
    & \multicolumn{1}{c}{}
    & \multicolumn{1}{c}{}
    & \multicolumn{1}{c}{}
    & \multicolumn{1}{c}{\textbf{Time}}
    & \multicolumn{1}{c}{\textbf{Space}}
    & \multicolumn{1}{c}{}
    & \multicolumn{1}{c}{}
    & \multicolumn{1}{c}{}
    & \multicolumn{1}{c}{}
    & \multicolumn{1}{c}{}
    \\
    \cmidrule(l){1-1} \cmidrule(l){2-2} \cmidrule(l){3-9} \cmidrule(l){10-16}
    \multicolumn{1}{l}{aes}
        & 23
        & \multicolumn{1}{r}{0.48}
        & \multicolumn{1}{r}{$\sim$0.01}
        & \multicolumn{1}{r}{342.11}
        & \multicolumn{1}{r}{-341.63}
        & 705.38
        & \multicolumn{1}{r}{16}
        & 14
        & \multicolumn{1}{r}{0.48}
        & \multicolumn{1}{r}{0.013}
        & \multicolumn{1}{r}{TO}
        & \multicolumn{1}{r}{-}
        & -
        & \multicolumn{1}{r}{16}
        & 14
        \\
    \multicolumn{1}{l}{backprop}
        & 34
        & \multicolumn{1}{r}{0.13}
        & \multicolumn{1}{r}{0.11}
        & \multicolumn{1}{r}{112.80}
        & \multicolumn{1}{r}{-112.56}
        & 470.00
        & \multicolumn{1}{r}{5}
        & 5
        & \multicolumn{1}{r}{0.14}
        & \multicolumn{1}{r}{0.024}
        & \multicolumn{1}{r}{TO}
        & \multicolumn{1}{r}{-}
        & -
        & \multicolumn{1}{r}{5}
        & 5
        \\
    \multicolumn{1}{l}{basicmath}
        & 21
        & \multicolumn{1}{r}{0.14}
        & \multicolumn{1}{r}{$\sim$0.01}
        & \multicolumn{1}{r}{102.83}
        & \multicolumn{1}{r}{-102.69}
        & 711.63
        & \multicolumn{1}{r}{7}
        & 7
        & \multicolumn{1}{r}{0.19}
        & \multicolumn{1}{r}{0.086}
        & \multicolumn{1}{r}{1362.58}
        & \multicolumn{1}{r}{-1362.30}
        & 4936.88
        & \multicolumn{1}{r}{7}
        & 7
        \\
    \multicolumn{1}{l}{bitcount}
        & 7
        & \multicolumn{1}{r}{0.039}
        & \multicolumn{1}{r}{$\sim$0.01}
        & \multicolumn{1}{r}{14.73}
        & \multicolumn{1}{r}{-14.69}
        & 371.97
        & \multicolumn{1}{r}{3}
        & 3
        & \multicolumn{1}{r}{0.062}
        & \multicolumn{1}{r}{$\sim$0.01}
        & \multicolumn{1}{r}{223.88}
        & \multicolumn{1}{r}{-223.82}
        & 3492.67
        & \multicolumn{1}{r}{3}
        & 3
        \\
    \multicolumn{1}{l}{cfd}
        & 51
        & \multicolumn{1}{r}{0.12}
        & \multicolumn{1}{r}{TO}
        & \multicolumn{1}{r}{TO}
        & \multicolumn{1}{r}{-}
        & -
        & \multicolumn{1}{r}{-}
        & 2
        & \multicolumn{1}{r}{0.14}
        & \multicolumn{1}{r}{TO}
        & \multicolumn{1}{r}{TO}
        & \multicolumn{1}{r}{-}
        & -
        & \multicolumn{1}{r}{-}
        & 2
        \\
    \multicolumn{1}{l}{crc32}
        & 24
        & \multicolumn{1}{r}{0.31}
        & \multicolumn{1}{r}{$\sim$0.01}
        & \multicolumn{1}{r}{262.82}
        & \multicolumn{1}{r}{-262.51}
        & 834.88
        & \multicolumn{1}{r}{11}
        & 8
        & \multicolumn{1}{r}{0.33}
        & \multicolumn{1}{r}{0.012}
        & \multicolumn{1}{r}{3867.11}
        & \multicolumn{1}{r}{-3866.77}
        & 11307.34
        & \multicolumn{1}{r}{11}
        & 8
        \\
    \multicolumn{1}{l}{fft}
        & 20
        & \multicolumn{1}{r}{0.14}
        & \multicolumn{1}{r}{$\sim$0.01}
        & \multicolumn{1}{r}{101.34}
        & \multicolumn{1}{r}{-101.20}
        & 711.66
        & \multicolumn{1}{r}{7}
        & 7
        & \multicolumn{1}{r}{0.23}
        & \multicolumn{1}{r}{$\sim$0.01}
        & \multicolumn{1}{r}{1485.63}
        & \multicolumn{1}{r}{-1485.39}
        & 6289.71
        & \multicolumn{1}{r}{7}
        & 7
        \\
    \multicolumn{1}{l}{gsm}
        & 24
        & \multicolumn{1}{r}{0.11}
        & \multicolumn{1}{r}{$\sim$0.01}
        & \multicolumn{1}{r}{191.03}
        & \multicolumn{1}{r}{-190.91}
        & 1603.95
        & \multicolumn{1}{r}{5}
        & 4
        & \multicolumn{1}{r}{0.14}
        & \multicolumn{1}{r}{$\sim$0.01}
        & \multicolumn{1}{r}{2799.07}
        & \multicolumn{1}{r}{-2798.71}
        & 18722.88
        & \multicolumn{1}{r}{5}
        & 4
        \\
    \multicolumn{1}{l}{heartwall}
        & 35
        & \multicolumn{1}{r}{0.17}
        & \multicolumn{1}{r}{$\sim$0.01}
        & \multicolumn{1}{r}{571.87}
        & \multicolumn{1}{r}{-571.69}
        & 3124.97
        & \multicolumn{1}{r}{3}
        & 3
        & \multicolumn{1}{r}{0.28}
        & \multicolumn{1}{r}{$\sim$0.01}
        & \multicolumn{1}{r}{TO}
        & \multicolumn{1}{r}{-}
        & -
        & \multicolumn{1}{r}{3}
        & 3
        \\
    \multicolumn{1}{l}{hotspot3D}
        & 57
        & \multicolumn{1}{r}{0.71}
        & \multicolumn{1}{r}{TO}
        & \multicolumn{1}{r}{TO}
        & \multicolumn{1}{r}{-}
        & -
        & \multicolumn{1}{r}{-}
        & 2
        & \multicolumn{1}{r}{0.83}
        & \multicolumn{1}{r}{TO}
        & \multicolumn{1}{r}{TO}
        & \multicolumn{1}{r}{-}
        & -
        & \multicolumn{1}{r}{-}
        & 2
        \\
    \multicolumn{1}{l}{lud}
        & 26
        & \multicolumn{1}{r}{0.08}
        & \multicolumn{1}{r}{$\sim$0.01}
        & \multicolumn{1}{r}{89.75}
        & \multicolumn{1}{r}{-89.66}
        & 1048.48
        & \multicolumn{1}{r}{3}
        & 3
        & \multicolumn{1}{r}{0.086}
        & \multicolumn{1}{r}{$\sim$0.01}
        & \multicolumn{1}{r}{1321.66}
        & \multicolumn{1}{r}{-1321.56}
        & 13216.60
        & \multicolumn{1}{r}{3}
        & 3
        \\
    \multicolumn{1}{l}{nw}
        & 33
        & \multicolumn{1}{r}{0.06}
        & \multicolumn{1}{r}{10.25}
        & \multicolumn{1}{r}{61.55}
        & \multicolumn{1}{r}{-51.23}
        & 5.97
        & \multicolumn{1}{r}{2}
        & 2
        & \multicolumn{1}{r}{0.068}
        & \multicolumn{1}{r}{0.15}
        & \multicolumn{1}{r}{981.69}
        & \multicolumn{1}{r}{-981.47}
        & 4503.17
        & \multicolumn{1}{r}{2}
        & 2
        \\
    \multicolumn{1}{l}{particlefilter}
        & 38
        & \multicolumn{1}{r}{0.37}
        & \multicolumn{1}{r}{70.34}
        & \multicolumn{1}{r}{451.48}
        & \multicolumn{1}{r}{-380.77}
        & 6.38
        & \multicolumn{1}{r}{9}
        & 9
        & \multicolumn{1}{r}{0.37}
        & \multicolumn{1}{r}{141.54}
        & \multicolumn{1}{r}{TO}
        & \multicolumn{1}{r}{-}
        & -
        & \multicolumn{1}{r}{9}
        & 9
        \\
    \multicolumn{1}{l}{sha1}
        & 21
        & \multicolumn{1}{r}{0.14}
        & \multicolumn{1}{r}{0.03}
        & \multicolumn{1}{r}{195.86}
        & \multicolumn{1}{r}{-195.69}
        & 1119.20
        & \multicolumn{1}{r}{4}
        & 2
        & \multicolumn{1}{r}{0.12}
        & \multicolumn{1}{r}{0.036}
        & \multicolumn{1}{r}{TO}
        & \multicolumn{1}{r}{-}
        & -
        & \multicolumn{1}{r}{4}
        & 2
        \\
    \multicolumn{1}{l}{sha2}
        & 25
        & \multicolumn{1}{r}{0.17}
        & \multicolumn{1}{r}{10.21}
        & \multicolumn{1}{r}{107.51}
        & \multicolumn{1}{r}{-97.13}
        & 10.36
        & \multicolumn{1}{r}{7}
        & 7
        & \multicolumn{1}{r}{0.17}
        & \multicolumn{1}{r}{2.02}
        & \multicolumn{1}{r}{1585.18}
        & \multicolumn{1}{r}{-1582.99}
        & 723.83
        & \multicolumn{1}{r}{7}
        & 7
        \\
    \multicolumn{1}{l}{stringsearch}
        & 28
        & \multicolumn{1}{r}{0.11}
        & \multicolumn{1}{r}{0.73}
        & \multicolumn{1}{r}{203.88}
        & \multicolumn{1}{r}{-203.04}
        & 242.71
        & \multicolumn{1}{r}{3}
        & 3
        & \multicolumn{1}{r}{0.11}
        & \multicolumn{1}{r}{0.61}
        & \multicolumn{1}{r}{3108.92}
        & \multicolumn{1}{r}{-3108.20}
        & 4317.94
        & \multicolumn{1}{r}{3}
        & 3
        \\
    \multicolumn{1}{l}{susan}
        & 21
        & \multicolumn{1}{r}{0.09}
        & \multicolumn{1}{r}{$\sim$0.01}
        & \multicolumn{1}{r}{213.63}
        & \multicolumn{1}{r}{-213.54}
        & 2350.17
        & \multicolumn{1}{r}{2}
        & 2
        & \multicolumn{1}{r}{0.09}
        & \multicolumn{1}{r}{$\sim$0.01}
        & \multicolumn{1}{r}{3314.91}
        & \multicolumn{1}{r}{-3314.82}
        & 35377.91
        & \multicolumn{1}{r}{2}
        & 2
        \\
    \cmidrule(l){1-1} \cmidrule(l){2-2} \cmidrule(l){3-9} \cmidrule(l){10-16}
    \multicolumn{1}{l}{Average}
        & -
        & \multicolumn{1}{r}{0.17}
        & \multicolumn{1}{r}{6.11}
        & \multicolumn{1}{r}{201.54}
        & \multicolumn{1}{r}{\textbf{-195.26}}
        & \textbf{887.84} 
        & \multicolumn{1}{r}{-}
        & -
        & \multicolumn{1}{r}{0.14}
        & \multicolumn{1}{r}{0.29}
        & \multicolumn{1}{r}{2006.06}
        & \multicolumn{1}{r}{\textbf{-2004.62}}
        & \textbf{10288.89} 
        & \multicolumn{1}{r}{-}
        & -
        \\
    \bottomrule
\end{tabular}
    \caption{Experimental results for four CGRAs; $\Delta$T and CTR (Compilation Time Ratio) are the difference and ratio of compilation time between the compared approaches. Each experiment had a 4000 seconds timeout. The average excludes benchmarks for which one of the tools had a timeout.}
    \label{tab:benchres}
\end{table*}

\subsubsection{Compilation time comparison}

Tab.~\ref{tab:benchres} also shows the compilation times obtained in our experiments. For all benchmarks, we observe a significant reduction in compilation time, with average speedups of $30.85\times$, $103.76\times$, $887.84\times$, and $10288.89\times$, for $2\times 2$, $5\times 5$, $10\times 10$ and $20\times 20$ CGRAs sizes. For an accurate comparison, we excluded from the average speedup calculation the cases for which one of the tools had a timeout. These results show how scalability is enhanced with our approach.
The speedup of $10288.89\times$ stands out, with~\cite{Tirelli2023} having to search across the entire mapping space, while our methodology benefits from its compositionality. 
By examining individual benchmarks, we find that our approach achieves a compilation speedup in almost every cases, with the exception of one case of a timeout and two cases of minimal difference.

Scalability is heavily influenced by the number of PEs in the grid, even for DFGs with as few as 23 nodes, e.g., aes benchmark.
Fig.~\ref{fig:aesmonosat} shows how CGRA size impacts each approach. While the compilation times of SAT-MapIt~\cite{Tirelli2023} increase with the CGRA size, the compilation times of our approach remain consistently small regardless of the grid size. This speedup comes from our decoupling of time and space, which enables independent searches over each dimension that are smaller and thus easier to navigate than the full mapping space. Furthermore, the search over the spatial dimension is accelerated by leveraging information obtained from the temporal solution.

\subsubsection{Limitations of our work}

Currently, our approach only targets architectures in which every PE can read the internal register of neighboring PEs. This makes the decoupling of space and time easier, but increases the complexity of the hardware design. Future work will focus on overcoming this restriction.


\section{Conclusion}

We propose a scalable CGRA mapping approach that effectively decouples the space and time dimensions and explores them in isolation. It first finds a time solution suitable for the target CGRA, by utilizing an SMT formulation and solver, and then explores the spatial dimension to find a space solution by utilizing a monomorphism search. Our experimental results showcase that our approach suffers no reduction in the quality of results achieved, finding mappings of quality similar to those found by state-of-the-art methods, while providing significantly better scalability, with an average speedup of $10288.89\times$ when compiling for a $20\times 20$ CGRA.


\bibliographystyle{IEEEtran}
\bibliography{references}

\end{document}